\documentclass[final]{aipproc}

\layoutstyle{6x9}

\usepackage{amsfonts,amssymb,amsmath,bm}
\usepackage{graphicx}

\newcommand{\be}{\begin{equation}}
\newcommand{\bea}{\begin{eqnarray}}
\newcommand{\ee}{\end{equation}}
\newcommand{\eea}{\end{eqnarray}}

\begin{document}

\title{The IR sector of QCD: lattice versus Schwinger-Dyson equations}


\author{Daniele Binosi}{
  address={European Centre for Theoretical Studies in Nuclear Physics and Related Areas (ECT*), Villa Tambosi, Strada delle Tabarelle 286, I-38050 Villazzano (TN), Italy}
}

\begin{abstract}
Important information about the infrared dynamics of QCD is encoded in the behavior of its (of-shell) Green's functions, most notably the gluon and the ghost propagators.  Due to recent improvements in the quality of lattice data and the truncation schemes employed for the Schwinger-Dyson equations we have now reached a point where the interplay between these two non-perturbative tools can be most fruitful. In this talk several of the above points will be reviewed, with particular emphasis on the implications for the ghost sector, the non-perturbative effective charge of QCD, and the Kugo-Ojima function.
\end{abstract}

\maketitle

%%%%%%%%%%%%%%%%%%%%%%%%%%%%%%%%%%%%%%%%%%%%
%% MAINMATTER
%%%%%%%%%%%%%%%%%%%%%%%%%%%%%%%%%%%%%%%%%%%%

\section{Introduction}

\noindent  Even though  the  Green's (correlation)  functions of  pure
Yang-Mills theories  are not {\it  per se} physical objects  (given that 
they depend explicitly on the gauge-fixing  and renormalization
scheme used),  the discovery and understanding of their infrared
(IR) properties  has become an  increasingly active topic.
In fact, the prevailing opinion to date is that they 
represent crucial pieces in our effort to unravel the non-perturbative 
QCD dynamics and to fully understand confinement.
 
The exploration of the IR sector of Yang-Mills theories is currently pursued
through mainly two non perturbative tools, namely the lattice -- where
space-time is discretized and the quantities of interest are evaluated
numerically --  and the Schwinger-Dyson equations  -- corresponding to
the infinite  set of integral  equation governing the dynamics  of the
Green's functions.

Recent   years   have   witnessed   a   lot  of   progress   in   both
methods~\cite{Cucchieri:2010xr,Binosi:2007pi},  and it  looks  like we
have  come to  a point  in which  it is  meaningful  to systematically
compare SDE results with lattice predictions.

For the particular case of the  gluon propagator and ghost dressing function, the solutions of the Landau gauge SDEs fall into two very distinct classes, namely:
\begin{itemize}
\item{\it Massive solutions}~\cite{Boucaud:2008ji,Aguilar:2008xm}: These solutions show a gluon propagator and a ghost dressing function that saturate in the IR to a finite (and non-zero) value, in complete agreement with 
the mechanism of confinement through thick vortex condensation, proposed by Cornwall~\cite{Cornwall:1981zr}.
\item {\it Scaling solutions}~\cite{Fischer:2006ub}: These solutions are characterized by power-law behavior with well-defined exponents, and lead to an IR vanishing gluon propagator  and, correspondingly, an IR diverging ghost dressing function; they are tailored to satisfy the confinement scenarios of  Kugo-Ojima~\cite{Kugo:1979gm} and Gribov-Zwanzinger~\cite{Gribov:1977wm}.
\end{itemize}

\begin{figure}[!t]
%\begin{center}
\includegraphics[scale=1]{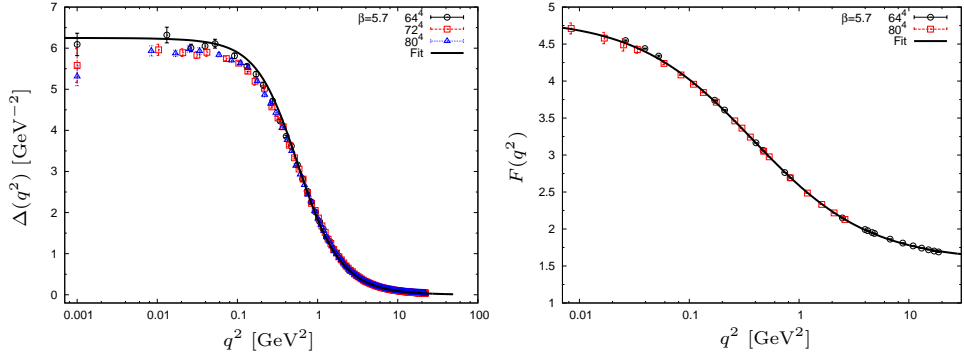}
\caption{Lattice data for the SU(3) gluon propagator and ghost dressing function~\cite{Bogolubsky:2007ud}. Solid lines corresponds to fits  in terms of a massive gluon propagator and a finite ghost dressing function.}
\label{fig1}
%\end{center}
\end{figure}

Over the last few years, ab-initio lattice computations
have been crucial in deciding which of these two types of solutions are in fact realized 
in (Landau gauge!) QCD. As a result,  a consistent and  clear picture has  emerged that
strongly advocates  for the massive solutions.   In Fig.~\ref{fig1} we
show     the     lattice     data     for     the     Landau     gauge
SU(3)~\cite{Bogolubsky:2007ud}  gluon  propagator  and ghost  dressing
function.  One can  see  the appearance  of  a plateau  for the  gluon
propagator in the deep IR region (which is one of the most salient and
distinctive predictions  of the the gluon  mass generation mechanism),
and no  enhancement for  the ghost dressing  function in the  deep IR,
where instead, it again saturates  to a constant. The lattice data can
be in  fact accurately fitted in  terms of a  massive gluon propagator
and a finite ghost  dressing function (see Fig.~\ref{fig1}); indeed this is
also valid for all lattice simulations available from different groups
(e.g., for the SU(2) case~\cite{Cucchieri:2010xr,Cucchieri:2007md}).

Landau gauge lattice studies therefore seem to rule out the possibility of scaling solutions with nontrivial infrared exponents (and consequently the Kugo-Ojima scenario). It also disfavor the original formulation of the Gribov-Zwanziger scenario, though the drastic modifications brought in by the inclusion of dimension two condensates, reconcile it with the lattice results~\cite{Dudal:2008sp}.

In what  follows we will concentrate on the  PT-BFM
(pinch       technique      -      background       field      method)
framework~\cite{Cornwall:1981zr,Cornwall:1989gv}, where the aforementioned lattice findings 
may be naturally accommodated. In fact, the discovery of the key underlying ingredient, namely 
the dynamical   generation   of  a   gluon
mass, coincided historically with the invention of the  PT~\cite{Cornwall:1981zr}, 
long before any lattice simulations were even contemplated.

\section{PT-BFM equations and numerical results}

\begin{figure}[!t]
\includegraphics[scale=1.1]{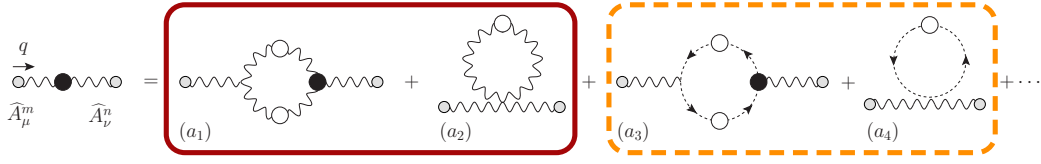}
\caption{The block-wise transverse SDE for the gluon self-energy in the  PT-BFM framework. The dots indicates that two-loop dressed gluon and ghost diagrams have been omitted.}
\label{fig0}
\end{figure}

{\bf PT-BFM equations}. The application of  the pinch technique~\cite{Cornwall:1989gv} to the conventional gluon self-energy SDE, gives rise {\it dynamically} to a new SDE~\cite{Binosi:2007pi} (see Fig.~\ref{fig0}), where (i) on the lhs the PT-BFM self-energy $\widehat{\Pi}_{\mu\nu}$ appears, whereas (ii) on the rhs the graphs display the conventional self-energy  $\Pi_{\mu\nu}$ but are made out from BFM vertices which satisfy simple Ward identities. This implies in turn that the new SDE is composed of one- and two-loop dressed gluon and ghost contributions which are individually transverse. 
The conventional and BFM self-energies are related through the identity~\cite{Grassi:1999tp} $\Delta=[1+G]^2\widehat{\Delta}$ where $G$ is on of the form factors of a certain auxiliary function $\Lambda_{\mu\nu}=g_{\mu\nu}G+(q_\mu q_\nu/q^2)L$. Then considering only the one-loop dressed diagrams of Fig.~\ref{fig0} and tracing out the transverse projectors, one can write
\be
\Delta^{-1}(q^2)=\frac{q^2+\mathrm{i}\sum_{i=1}^4(a_i)}{[1+G(q^2)]^2}.
\label{gSDE}
\ee
We next express the background three gluon and ghost vertex appearing in diagrams $(a_1)$ and $(a_3)$ as a function of the gluon and ghost self-energies in such a way as to (i) automatically satisfy the Ward identities and (ii) to introduce the massless poles necessary to trigger the Schwinger mechanism~\cite{Schwinger:1962tn}. The Ans\"atz we will use is
\be
\widetilde{\Gamma}_{\alpha\mu\nu}=\Gamma_{\alpha\mu\nu}^{(0)}+\mathrm{i}\frac{q_\alpha}{q^2}\left[\Pi_{\mu\nu}(k+q)-\Pi_{\mu\nu}(k)\right],
\ee
and similarly for the ghost vertex $\widetilde{\Gamma}_{\alpha}$. The resulting expression for the SDE~(\ref{gSDE}) after projection to the Landau gauge is very lengthy~\cite{Aguilar:2008xm}. The important point is that the resulting gluon self-energy is IR finite, $\Delta^{-1}(0)>0$.

\begin{figure}[!t]
\includegraphics[scale=0.88]{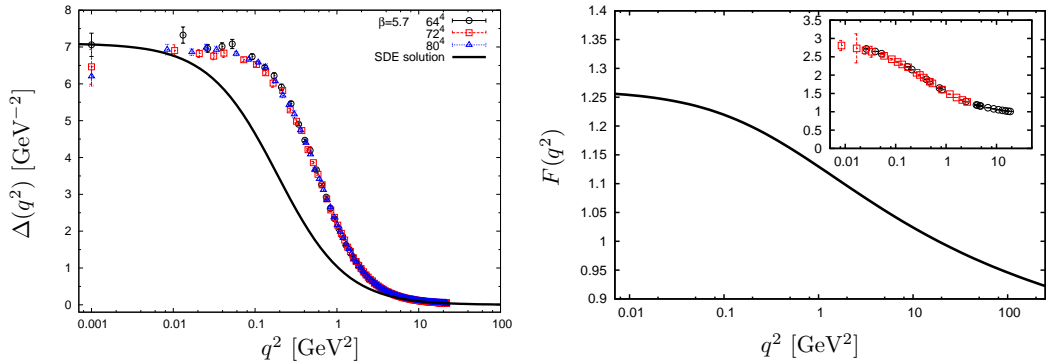}
\caption{Gluon propagator and ghost dressing function for the SU(3) gauge group, and comparison with the corresponding lattice data of ~\cite{Bogolubsky:2007ud}.}
\label{fig2}
\end{figure}

\noindent{\bf Gluon and ghost Green's functions}. The solutions of the
PT-BFM  equations~(\ref{gSDE})
SU(3)  case~\cite{Aguilar:2008xm} are  shown  in Fig.~\ref{fig2}.  The
agreement found between the SDE  and the lattice results allows one to
study other quantities of interest by using the lattice directly as an
input into the various SDE.  The general strategy adopted in this case
is the following.  One takes the lattice gluon  propagator as an input
for the ghost SDE; then solves for the ghost dressing function, tuning
the coupling  constant $g$ such  that the  solution gives the
best possible approximation to  the lattice result. Obviously one must
check that the coupling so obtained (at the renormalization scale used
for  the  computation) is  fully  consistent  with known  perturbative
results (obtained in  the MOM scheme, which is the  scheme used in our
computations), and this indeed what happens.
At this point the system is  ``tuned'', and one can construct and
analyze other  quantities that are built from $\Delta$, $F$ and  $g$, as
done in the next two sections.

\noindent{\bf The $G$ and $L$ auxiliary functions}. The first quantities that can be studied by means of the  procedure just described, are the auxiliary functions $G$ and $L$, which in the Landau gauge one can prove to be related to the dressing function $F(q^2)$ through the BRST identity~\cite{Kugo:1995km} $F^{-1}(q^2)=1+G(q^2)+L(q^2)$.
Since, under very general conditions on the gluon and ghost propagators, $L(0)=0$ one has the IR relation $F^{-1}(0)=1+G(0)$. Thus we see that a divergent (or {\it enhanced}) dressing function requires the condition $G(0)=-1$. The latter looks suspiciously similar to the Kugo-Ojima confinement criterion, demanding that a certain function $u(q^2)$ (the Kugo-Ojima function) acquires the IR value $u(0)=-1$. 
Indeed, it is possible to show (in the Landau gauge only!) 
that $G$ is nothing but the Kugo-Ojima function~\cite{Kugo:1995km}: $u(q^2)\equiv G(q^2)$.

\begin{figure}[!t]
\includegraphics[scale=0.86]{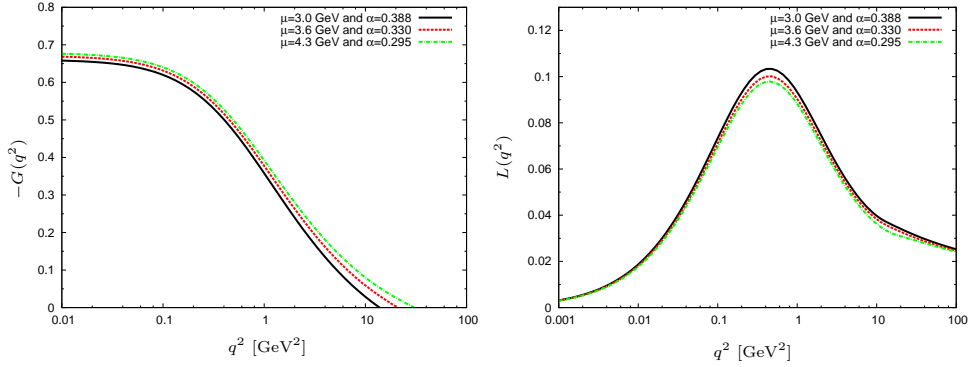}
\caption{The form factors $-G(q^2)$ and $L(q^2)$ determined from Eq. (4.2) at different normalization points $\mu$, using the procedure specified in the text.}
\label{fig3}
\end{figure}

In Fig.~(\ref{fig3}) we show these auxiliary functions calculated at different renormalization points. One can see that indeed $L(0)=0$ and that in general $L$ is suppressed with respect to $G$. In addition one finds that the function $G$ saturates at an IR value bigger than $-1$ (around $-2/3$ for the renormalization points chosen) once again excluding IR enhancement of the ghost dressing function. Also these results are in good agreement with direct lattice calculations of the Kugo-Ojima function. 

\noindent{\bf  The effective  charge}.  Another important  information
that can be extracted from the  PT-BFM equations is the running of the
QCD effective charge  for a wide
range of physical momenta, and,  in particular, its behavior and value
in  the  deep  IR.  The effective charge
is invariant under the renormalization group  (RG), and 
lies  at  the  interface  between
perturbative  and   non-perturbative  effects  in   QCD,  providing  a
continuous interpolation between  two physically distinct regimes: the
deep UV, where perturbation theory is reliable, and the deep IR, where
non-perturbative techniques must be employed.

There are two possible RG-invariant products  on which a definition of the effective charge can be based, namely $\widehat{r}(q^2)=g^2(\mu^2)\Delta(q^2)F^2(q^2)$, which exploits the non-renormalization property of the ghost vertex in the Landau gauge, and $\widehat{d}(q^2)=g^2(\mu^2)\widehat{\Delta}(q^2)$, which relies on the fact that in the background quantities  satisfy Ward (as opposed to Slavnov-Taylor) identities. 
These two {\it dimensionful} quantities [mass dimension of $-2$] share an important common ingredient, namely the scalar cofactor of the gluon propagator, $\Delta(q^2)$, which actually sets the scale. The next step is to extract a {\it dimensionless} quantity that would correspond to the non-perturbative effective charge. Perturbatively, i.e., for asymptotically large momenta, it is clear that the mass scale is saturated simply by $q^2$, the bare gluon propagator, and the effective charge is defined by pulling a $q^{-2}$ out of the corresponding RG-invariant quantity. Of course,  in the IR the gluon propagator becomes effectively massive; therefore, particular care is needed in deciding exactly what combination of mass scales ought to be pulled out. The correct procedure in such a case~\cite{Cornwall:1981zr} is to pull out a massive propagator of the form (in the Euclidean space) $[q^2+m^2(q^2)]^{-1}$, with $m^2(q^2)$ the dynamical gluon mass. One has~\cite{Aguilar:2008fh}
\be
\alpha_{\mathrm{gh}}(q^2)=\alpha(\mu^2)[q^2+m^2(q^2)]\Delta(q^2)F^2(q^2),\qquad 
\alpha(q^2)=\alpha(\mu^2)[q^2+m^2(q^2)]\widehat{\Delta}(q^2),
\label{echs}
\ee
and, due to the BRST identity between $F$, $G$ and $L$, the two effective charges are related:
\be
\alpha(q^2)=\alpha_{\mathrm{gh}}(q^2)\left[1+\frac{L(q^2)}{1+G(q^2)}\right]^2.
\ee
Since $L(0)=0$ we therefore see that not only the two effective charges coincide in the UV region where they should reproduce the perturbative result, but also in the deep IR where one has $\alpha(0)=\alpha_{\mathrm{gh}}(0)$. In addition, due to the relative suppression of $L$ as compared to $G$, 	even in the region of intermediate momenta, where the difference reaches its maximum, the relative difference between the two charges is small. 

\begin{figure}[!t]
\includegraphics[scale=1]{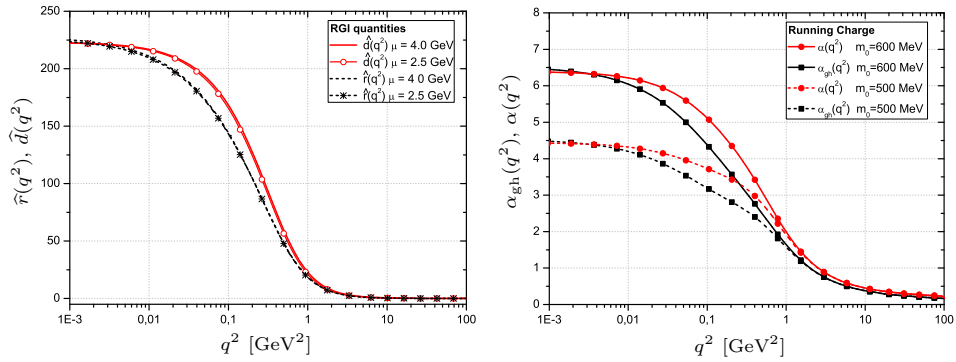}
\caption{\label{fig4}
{\it Left panel}: Comparison between the two RG-invariant products $\widehat{d}(q^2)$ (solid line) and $\widehat{r}(q^2)$ (dashed line). {\it Right panel}: Comparison between the QCD effective charge extracted from lattice data:  $\alpha(q^2)$ (red line with circles) 
and $\alpha_{\mathrm{gh}}$ (black line with squares) for two different masses $m_0$.}
\end{figure}

In Fig.~\ref{fig4} we show both a check of the RG-invariance of the combinations $\widehat{r}$ and $\widehat{d}$, as well as a comparison between the effective charges~(\ref{echs}) extracted from the lattice data for two different values of the running gluon mass. 

\section{Conclusions}

In this talk we have outlined the salient features of the SDEs formulated within the PT-BFM framework. 
A number of examples of the considerable potential offered by their interplay with the lattice simulations have also been given. Clearly, several aspects need to be further investigated, and in particular: (i) to improve the agreement between 
the PT-BFM and lattice results by devising better vertex Ans\"atze for implementing the Schwinger mechanism; (ii) to study on the lattice how the results change when calculations are performed in gauges other than the Landau; (iii) ideally, to code on the lattice the BFM in the Feynman gauge, along the lines suggested in~\cite{Dashen:1980vm}, where a plethora of results are known to be free from gauge artifacts~\cite{Cornwall:1989gv}.

\bibliographystyle{aipproc}   % if natbib is available

\begin{thebibliography}{9}

%\cite{Cucchieri:2010xr}
\bibitem{Cucchieri:2010xr}
 For a thorough and concise review of the recent lattice literature on the topic see  A.~Cucchieri and T.~Mendes,
  %``Numerical test of the Gribov-Zwanziger scenario in Landau gauge,''
  PoS {\bf QCD-TNT09}, 026 (2009)
  %[arXiv:1001.2584 [hep-lat]].
  %%CITATION = POSCI,QCD-TNT09,026;%%

%\cite{Aguilar:2006gr}
%\bibitem{Aguilar:2006gr}
  %A.~C.~Aguilar and J.~Papavassiliou,
  %``Gluon mass generation in the PT-BFM scheme,''
  %JHEP {\bf 0612}, 012 (2006);
  %[arXiv:hep-ph/0610040].
  %%CITATION = JHEPA,0612,012;%%  
   %\cite{Binosi:2007pi}
\bibitem{Binosi:2007pi}
  D.~Binosi and J.~Papavassiliou,
  %``Gauge-invariant truncation scheme for the Schwinger-Dyson equations of
  %QCD,''
  Phys.\ Rev.\  D {\bf 77}, 061702(R) (2008);
  %[arXiv:0712.2707 [hep-ph]].
  %%CITATION = PHRVA,D77,061702;%%
%\cite{Binosi:2008qk}
%\bibitem{Binosi:2008qk}
  %D.~Binosi and J.~Papavassiliou,
  %``New Schwinger-Dyson equations for non-Abelian gauge theories,''
  JHEP {\bf 0811}, 063 (2008).
 % [arXiv:0805.3994 [hep-ph]].
  %%CITATION = JHEPA,0811,063;%%

%\cite{Boucaud:2008ji}
\bibitem{Boucaud:2008ji}
  P.~Boucaud, J.~P.~Leroy, A.~L.~Yaouanc, J.~Micheli, O.~Pene and J.~Rodriguez-Quintero,
  %``IR finiteness of the ghost dressing function from numerical resolution of
  %the ghost SD equation,''
  JHEP {\bf 0806}, 012 (2008).
  %[arXiv:0801.2721 [hep-ph]].
  %%CITATION = JHEPA,0806,012;%%  

%\cite{Aguilar:2008xm}
\bibitem{Aguilar:2008xm}
  A.~C.~Aguilar, D.~Binosi and J.~Papavassiliou,
  %``Gluon and ghost propagators in the Landau gauge: Deriving lattice results
  %from Schwinger-Dyson equations,''
  Phys.\ Rev.\  D {\bf 78}, 025010 (2008).
  %[arXiv:0802.1870 [hep-ph]].
  %%CITATION = PHRVA,D78,025010;%%


%\cite{Cornwall:1981zr}
\bibitem{Cornwall:1981zr}
  J.~M.~Cornwall,
  %``Dynamical Mass Generation In Continuum QCD,''
  Phys.\ Rev.\  D {\bf 26}, 1453 (1982).
  %%CITATION = PHRVA,D26,1453;%%


%\cite{Fischer:2006ub}
\bibitem{Fischer:2006ub}
  See C.~S.~Fischer,
  %``Infrared properties of QCD from Dyson-Schwinger equations,''
  J.\ Phys.\ G {\bf 32}, R253 (2006), and references therein.
 % [arXiv:hep-ph/0605173].
  %%CITATION = JPHGB,G32,R253;%%

%\cite{Kugo:1979gm}
\bibitem{Kugo:1979gm}
  T.~Kugo and I.~Ojima,
  %``Local Covariant Operator Formalism Of Nonabelian Gauge Theories And Quark
  %Confinement Problem,''
  Prog.\ Theor.\ Phys.\ Suppl.\  {\bf 66}, 1 (1979).
  %%CITATION = PTPSA,66,1;%%

%\cite{Gribov:1977wm}
\bibitem{Gribov:1977wm}
  V.~N.~Gribov,
  %``Quantization of non-Abelian gauge theories,''
  Nucl.\ Phys.\  B {\bf 139}, 1 (1978);
  %%CITATION = NUPHA,B139,1;%%
%\cite{Zwanziger:1991gz}
%\bibitem{Zwanziger:1991gz}
  D.~Zwanziger,
  %``Vanishing of zero momentum lattice gluon propagator and color
  %confinement,''
  Nucl.\ Phys.\  B {\bf 364} (1991) 127.
  %%CITATION = NUPHA,B364,127;%%
%\cite{Zwanziger:1993dh}
%\bibitem{Zwanziger:1993dh}
  %D.~Zwanziger,
  %``Fundamental modular region, Boltzmann factor and area law in lattice gauge
  %theory,''
%  Nucl.\ Phys.\  B {\bf 412}, 657 (1994).
  %%CITATION = NUPHA,B412,657;%%

  %\cite{Bogolubsky:2007ud} 
\bibitem{Bogolubsky:2007ud}
  I.~L.~Bogolubsky, E.~M.~Ilgenfritz, M.~Muller-Preussker and A.~Sternbeck,
  %``The Landau gauge gluon and ghost propagators in 4D SU(3) gluodynamics in
  %large lattice volumes,''
  PoS {\bf LAT2007}, 290 (2007).
  %[arXiv:0710.1968 [hep-lat]].
  %%CITATION = POSCI,LAT2007,290;%%   



%\cite{Cucchieri:2007md}
\bibitem{Cucchieri:2007md}
  A.~Cucchieri and T.~Mendes,
  %``What's up with IR gluon and ghost propagators in Landau gauge? A puzzling
  %answer from huge lattices,''
  PoS {\bf LAT2007}, 297 (2007).
  %[arXiv:0710.0412 [hep-lat]].
  %%CITATION = POSCI,LAT2007,297;%%
  

%\cite{Dudal:2008sp}
\bibitem{Dudal:2008sp}
  D.~Dudal, J.~A.~Gracey, S.~P.~Sorella, N.~Vandersickel and H.~Verschelde,
  %``A refinement of the Gribov-Zwanziger approach in the Landau gauge: infrared
  %propagators in harmony with the lattice results,''
  Phys.\ Rev.\  D {\bf 78}, 065047 (2008).
 % [arXiv:0806.4348 [hep-th]].
  %%CITATION = PHRVA,D78,065047;%%



  

%%\cite{Farhi:1982vt}
%\bibitem{Farhi:1982vt}
%  E.~Farhi and R.~Jackiw,
%  %``Dynamical Gauge Symmetry Breaking. A Collection Of Reprints,''
%%\href{http://www.slac.stanford.edu/spires/find/hep/www?irn=1110667}{SPIRES entry}
%{\it  Singapore, Singapore: World Scientific ( 1982) 403p}.

%\cite{Cornwall:1989gv}
\bibitem{Cornwall:1989gv}
  J.~M.~Cornwall and J.~Papavassiliou,
  %``Gauge Invariant Three Gluon Vertex in QCD,''
  Phys.\ Rev.\  D {\bf 40}, 3474 (1989);
  %%CITATION = PHRVA,D40,3474;%%
%\cite{Binosi:2002ft}
%\bibitem{Binosi:2002ft}
  D.~Binosi and J.~Papavassiliou,
  %``The pinch technique to all orders,''
  Phys.\ Rev.\  D {\bf 66}(R), 111901 (2002);
  %[arXiv:hep-ph/0208189].
  %%CITATION = PHRVA,D66,111901;%%
%\cite{Binosi:2003rr}
%\bibitem{Binosi:2003rr}
 % D.~Binosi and J.~Papavassiliou,
  %``Pinch technique self-energies and vertices to all orders in perturbation
  %theory,''
  J.\ Phys.\ G {\bf 30}, 203 (2004);
  %[arXiv:hep-ph/0301096].
  %%CITATION = JPHGB,G30,203;%%
%\cite{Binosi:2009qm}
%\bibitem{Binosi:2009qm}
  D.~Binosi and J.~Papavassiliou,
  %``Pinch Technique: Theory and Applications,''
  Phys.\ Rept.\  {\bf 479}, 1 (2009).
  %[arXiv:0909.2536 [hep-ph]].
  %%CITATION = PRPLC,479,1;%%

%\cite{Grassi:1999tp}
\bibitem{Grassi:1999tp}
  P.~A.~Grassi, T.~Hurth and M.~Steinhauser,
  %``Practical algebraic renormalization,''
  Annals Phys.\  {\bf 288}, 197 (2001);
  %[arXiv:hep-ph/9907426].
  %%CITATION = APNYA,288,197;%%
%\cite{Binosi:2002ez}
%\bibitem{Binosi:2002ez}
  D.~Binosi and J.~Papavassiliou,
  %``Pinch technique and the Batalin-Vilkovisky formalism,''
  Phys.\ Rev.\  D {\bf 66}, 025024 (2002).
  %[arXiv:hep-ph/0204128].
  %%CITATION = PHRVA,D66,025024;%%
  
 %\cite{Schwinger:1962tn}
\bibitem{Schwinger:1962tn}
  J.~S.~Schwinger,
  %``Gauge Invariance and Mass,''
  Phys.\ Rev.\  {\bf 125}, 397-398 (1962);
%\cite{Schwinger:1962tp}
%\bibitem{Schwinger:1962tp}
  %J.~S.~Schwinger,
  %``Gauge Invariance and Mass. 2.,''
  Phys.\ Rev.\  {\bf 128}, 2425-2429 (1962).

%\cite{Kugo:1995km}
\bibitem{Kugo:1995km}
  T.~Kugo,
  %``The universal renormalization factors Z(1) / Z(3) and color confinement
  %condition in non-Abelian gauge theory,''
  arXiv:hep-th/9511033;
  %\cite{Grassi:2004yq}
%\bibitem{Grassi:2004yq}
  P.~A.~Grassi, T.~Hurth and A.~Quadri,
  %``On the Landau background gauge fixing and the IR properties of YM Green
  %functions,''
  Phys.\ Rev.\  D {\bf 70}, 105014 (2004);
% [arXiv:hep-th/0405104].
  %%CITATION = PHRVA,D70,105014;%%
  %%CITATION = HEP-TH/9511033;%%
%\cite{Aguilar:2009pp}
%\bibitem{Aguilar:2009pp}
  A.~C.~Aguilar, D.~Binosi and J.~Papavassiliou,
  %``Indirect determination of the Kugo-Ojima function from lattice data,''
  JHEP {\bf 0911}, 066 (2009).
  %[arXiv:0907.0153 [hep-ph]].
  %%CITATION = JHEPA,0911,066;%%

%\cite{Aguilar:2008fh}
\bibitem{Aguilar:2008fh}
  A.~C.~Aguilar, D.~Binosi and J.~Papavassiliou,
  %``Infrared finite effective charge of QCD,''
  PoS {\bf LC2008}, 050 (2008);
 % [arXiv:0810.2333 [hep-ph]].
  %%CITATION = POSCI,LC2008,050;%%
%\cite{Aguilar:2010gm}
%\bibitem{Aguilar:2010gm}
 % A.~C.~Aguilar, D.~Binosi and J.~Papavassiliou,
  %``QCD effective charges from lattice data,''
  JHEP {\bf 1007}, 002 (2010);
 % [arXiv:1004.1105 [hep-ph]].
  %%CITATION = JHEPA,1007,002;%%
%\cite{Aguilar:2009nf}
%\bibitem{Aguilar:2009nf}
  A.~C.~Aguilar, D.~Binosi, J.~Papavassiliou and J.~Rodriguez-Quintero,
  %``Non-perturbative comparison of QCD effective charges,''
  Phys.\ Rev.\  D {\bf 80}, 085018 (2009).
 % [arXiv:0906.2633 [hep-ph]].
  %%CITATION = PHRVA,D80,085018;%%

%\cite{Dashen:1980vm}
\bibitem{Dashen:1980vm}
  R.~F.~Dashen and D.~J.~Gross,
  %``The Relationship Between Lattice And Continuum Definitions Of The Gauge
  %Theory Coupling,''
  Phys.\ Rev.\  D {\bf 23}, 2340 (1981).
  %%CITATION = PHRVA,D23,2340;%%


\end{thebibliography}

\end{document}